# milq – Quantum Physics in Secondary School

Rainer Müller[1[0000-0003-1013-411X]] and Oxana Mishina[1,2[0000-0001-7546-6719]]

[1] Physikdidaktik, TU Braunschweig, Bienroder Weg 82, 38106 Braunschweig, Germany
[2] Dipartimento di Fisica, Università di Trieste, Via Valerio 2, 34127 Trieste, Italy

**Abstract.** The milq approach to quantum physics for high schools focuses on the conceptual questions of quantum physics. Students should be given the opportunity to engage with the world view of modern physics. The aim is to achieve a conceptually clear formulation of quantum physics with a minimum of formulas. In order to provide students with verbal tools they can use in discussions and argumentations we formulated four "reasoning tools". They help to facilitate qualitative discussions of quantum physics, allow students to predict quantum mechanical effects and help to avoid learning difficulties. They form a "beginners' axiomatic system" for quantum physics.

## 1  Introduction

Without doubt, quantum physics is one of the more difficult areas for teaching at high school level. Why is this so? When trying to pinpoint the difficulties more precisely, one finds that the explanation is not easy: (1) Quantum physics is experimentally poorly accessible with high school resources? Yes it is, but that is probably not the real problem. (2) Quantum physics is so difficult to teach because it is a mystery and nobody understands it? Popular accounts often give the impression that essential parts of quantum physics are conceptually poorly understood, especially in those areas (such as the measurement process) that are particularly interesting for discussion at high school. We will show in this article that, contrary to this belief, it is possible to discuss even the most difficult aspects of quantum physics at high school level if only the appropriate terms and concepts are used. (3) It is difficult to teach quantum physics at high school level because it is mathematically so complicated. This is not correct either. Consider the case of electricity: The Maxwell equations are mathematically much more complicated than the Schrödinger equation. Nevertheless, in the case of electricity, it has been possible to formulate the content in such a way that the basics of electricity are routinely taught in high school today.

The comparison between quantum physics and electricity is helpful to understand in which direction quantum physics education could be successfully developed. In electricity, it has been possible to identify some basic features for teaching which are regarded as "essential" for understanding. The corresponding terms have become so much a part of our everyday life that we take them for granted: Nobody would doubt that a proper introduction to electricity would include the concepts of charges and static electricity, current and voltage, stationary circuits, magnetic effects and induction and electromagnetic waves. In quantum physics, such a clarification process about basic and essential concepts has apparently never taken place. The curriculum is often based on the history of discovery – an approach that certainly would not work well in the case of electricity.

The aim of the milq concept has been to provide such a clarification of basic concepts in quantum physics for high school. The aim was not to make small corrections to existing concepts, but to restructure the subject matter for teaching from a different perspective. The German version of the course has been online since 2001 and is continuously developing. Since 2018, an English version is available and it will continue to grow (milq.info/en/).

In Germany, quantum physics has been an established part of the school's physics curriculum for several decades. There is a considerable amount of experience and research on teaching and learning quantum physics in secondary schools. The first curriculum explicitly based on the investigation of students' conceptions and learning difficulties was that of Fischler and Lichtfeldt [1]. They aimed at a "minimal conception" that strongly emphasizes the difference between classical and quantum phenomena. References to classical physics (Bohr's atomic model) are largely avoided. Already earlier, Brachner and Fichtner [2,3] had proposed a curriculum which, following the "Feynman Lectures" [4], focuses on the concept of probability amplitude. The "quantum mechanical fundamental principle" (see below) is explained using the example of the Mach-Zehnder interferometer. Several curricula have been developed based on Feynman's "arrow formalism" [5], e.g. the proposals of Bader [6], Küblbeck [7] or Werner [8]. In the English-speaking world, the "Visual Quantum Mechanics" concept by Zollman [9] can best be compared with these curricula in terms of their physical and mathematical requirement level. Simulation programs are used here, for example, to examine the spectra of LEDs and trace them back to the band structure in solids. Also the approach by Michelini et al. [10], where the polarization degrees of photons are related to states and operators, addresses a similar audience.

## 2  The milq concept

The milq approach has been developed since the mid-1990s in Munich and later in Braunschweig [11, 12]. It focuses on the conceptual questions of quantum physics. Students should be given the opportunity to engage with the world view of modern physics. It is not intended to strive for continuity with the notions of classical physics. Instead, especially those aspects of quantum physics are emphasized that involve a radical break with the classical concepts. The most prominent examples are Born's probability interpretation, which implies the departure from classical determinism, and superposition states in which quantum objects do not possess classically well-defined properties like position, trajectory, or energy.

Aspects of proper language play a major role in the milq approach, because many misconceptions and learning difficulties in quantum physics are related to language. Our language has developed in dealing with the phenomena of classical physics. It is only partially suitable for describing the completely different world of quantum phenomena. In university physics, mathematical formalism is the medium for communication about quantum physics. Language can play a subordinate role, because the clarifying reference to formalism is always possible in case of doubt. Linguistic inaccuracies, laboratory jargon and simplifying abbreviations are commonplace among physicists without causing much damage. This is different in high school physics. Here we do not have access to mathematical formalism so that we have to rely on language to communicate about quantum physics. For this reason, great emphasis was placed on the conceptual side of quantum physics: Conceptual clarity was the main focus in the development of the milq approach. It was considered important to provide concise terms that help us to speak systematically about quantum phenomena. One example of a helpful notion is the term "preparation", which describes the state of quantum objects operationally and in qualitative terms. It is thus the equivalent of the wave function in mathematical formalism, because the wave function is used to mathematically describe quantum objects that have undergone a certain preparation process.

## 3    Reasoning tools

Within the milq approach, we formulate a set of four qualitative rules that are supposed to contain the basic traits of quantum physics. They are called reasoning tools [13]. They can serve as a support in qualitative discussions about quantum phenomena. They enable students to predict certain kinds of quantum behavior and help to avoid learning difficulties. In this sense, the reasoning tools may be regarded as a kind of  "qualitative mini-axiomatic" for quantum physics. The four rules are as follows:

**Rule 1: Statistical behavior.**
Single events are not predictable, they are random. Only statistical predictions (for many repetitions) are possible in quantum physics.

**Rule 2: Interference of single quantum objects.**
Interference occurs if there are two or more "paths" leading to the same experimental result. Even if these alternatives are mutually exclusive in classical physics, none of them will be "realized" in a classical sense.

**Rule 3: Unique measurement results.**
Even if quantum objects in a superposition state need not have a fixed value of the measured quantity, one always finds a unique result upon measurement.

**Rule 4: Complementarity.**
Exemplary formulations are: "Which-path information and interference pattern are mutually exclusive" or "Quantum objects cannot be prepared for position and momentum simultaneously."

At the time when the reasoning tools were formulated, the concept of entanglement seemed too far away from high school physics to be taken into account. From today's perspective, entanglement appears to be an essential feature of quantum physics, especially under the increasingly important aspect of quantum information. If future applications in quantum technology will yield more experience what can be "done" with entanglement, it should possibly be added. One should add, however, that even in the mathematical theory of quantum mechanics, entanglement is not an independent axiom but a consequence of the principle of superposition.

## 4 The physical background of the reasoning tools

### 4.1 Rule 1: The probabilistic nature of quantum physics

Rule 1 contains Born's probability interpretation in qualitative form. The probabilistic character of quantum physics may be regarded as the one essential difference to classical physics. Classical physics is deterministic. If you throw a basketball at the right speed and angle, you can be sure that it will hit the basket – and this is repeatable. On the contrary, in the double-slit experiment with single electrons, the location where the next electron is detected cannot be predicted – nor can it be controlled experimentally. This is a general rule: In quantum physics, individual events cannot generally be predicted. This inability to predict is not based on subjective ignorance. The experiments on Bell's inequality show that there are no local hidden variables that could govern the outcome of an experiment. According to quantum physics, there is true randomness in nature.

However, quantum physics is not completely without laws. Its laws are statistical in nature. Many repetitions of the same experiment result in a distribution of measured values that – except for statistical deviations – is predictable and reproducible. In addition, there are certain states (eigenstates of observables) in which the outcome of experiments can be predicted with a probability of 1. The success of quantum physics is based on these regularities, making it the best tested theory in physics.

### 4.2 Rule 2: Interference and superposition states

Rule 2 is closely linked to a rule that was emphasized by Feynman [4]. It was called the "fundamental principle" by Brachner and Fichtner [2]. It reads as follows: "If there are different possibilities (paths) for the occurrence of a particular event and if the experimental setup does not determine uniquely that only one specific possibility was chosen, interference always occurs. If, on the other hand, each event in the experimental setup leaves a certain trace that can be used to decide which of the various possibilities was chosen, then interference will never occur."

A paradigmatic application of the fundamental principle is the quantum mechanical double slit experiment. If, with a suitable measuring device, it can be decided through which of the slits an electron has passed, no interference occurs. If, on the other hand, there is no means for assigning a specific slit to the electrons, they form an interference pattern on the screen.

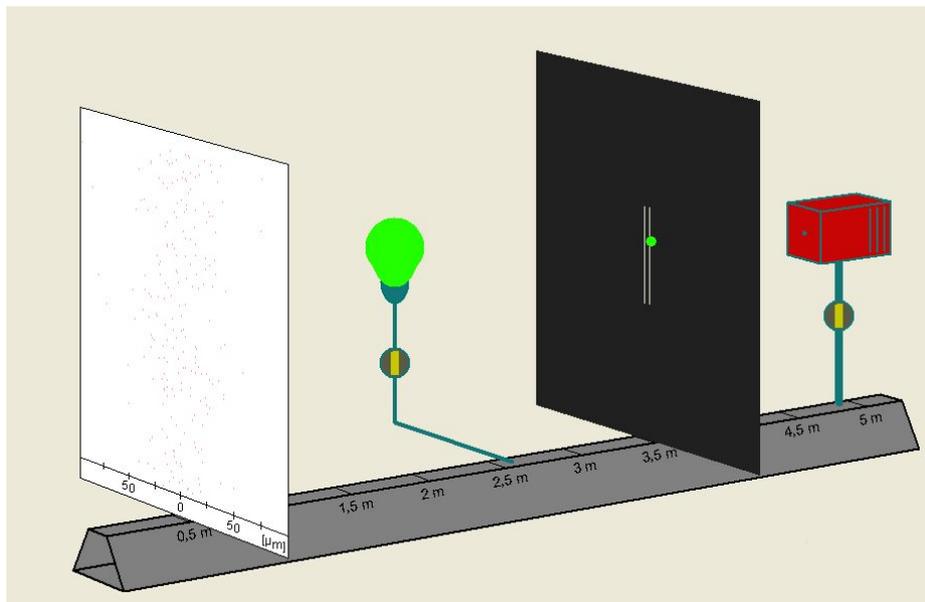

**Fig. 1.** Screenshot of the milq double slit simulation software

The double slit experiment can also be used to illustrate the statement: "In a superposition state, none of the conceivable alternatives will be realized in a classical sense." Within the double slit apparatus, the electrons are in a quantum mechanical superposition state in which the property "position" cannot be attributed to them. The electrons do not go "either left or right". This assumption would be incompatible with the experimental observation of double slit interference. Instead, they are in a superposition of both alternatives. Neither of the two classical alternatives (left or right) is actually realized. In quantum physics, electrons can no longer be regarded as objects with a definite position or trajectory. This fundamental insight has important consequences for our concept of atoms where it is crucial that electrons are delocalized objects.

### 4.3     **Rule 3: The measurement postulate**

Rule 3 contains the measurement postulate of quantum mechanics. Even with quantum objects in superposition states, measurements yield definite results. The result of a measurement is always one of the eigenvalues of the measured observable.

Following Feynman [4], this can be illustrated in the double-slit experiment by placing a light source behind the slits to illuminate the electrons passing by (figure 1). The electrons scatter some of the light and the scattered light is registered by a detector or the eye of an observer. In this way, a measurement of the observable "position" is performed. It turns out that for each individual electron, a flash of light is seen at a specific position behind one of the slits (visible in figure 1). This means: In the position measurement, each electron is found at a well-defined position – even if the electrons were in a state before the measurement in which they did not have the property "position". A similar position measurement is made at the detection screen: Each electron is found at a well-defined spot on the screen, even if it was in a delocalized state before the detection.

### 4.4     **Rule 4: Complementarity**

The Feynman lamp can also be used to illustrate rule 4. When the lamp is switched on and registers the path of each electron, no double-slit interference pattern emerges on the screen. Instead, a distribution without internal structure appears. Which-path information and interference pattern cannot be realized simultaneously, but are mutually exclusive. This is an example of two complementary quantities, such as location and momentum in Heisenberg's uncertainty relation.

### 4.5     **Epistemological status of the reasoning tools**

One advantage of the reasoning tools is particularly noteworthy: they are strictly correct. Nothing that is stated in the four rules has ever to be taken back, even if one advances to university level and to the most complicated facts of quantum mechanics. This distinguishes them from many models of school physics or chemistry, which have to be "unlearned" in the course of further studies and replaced by successor models (for example in the field of atomic physics).

The reasoning tools also claim to contain the qualitative key aspects of quantum physics quite completely. Of course, they lack anything that can be only expressed within the framework of mathematical formalism. In an epistemological sense, however, nothing entirely new is added, no matter how deep one goes.

## 5 Key experiments

In the milq approach, the reasoning tools are introduced by means of key experiments in which they become particularly evident. The aim is to give students the opportunity to test the application of the rules in real life situations. Two key experiments are discussed in detail. Simulation software has been written for both experiments. It is provided for free at the milq web page milq.tu-bs.de:

1. The first key experiment is the double-slit experiment with single quantum objects, which according to Feynman [4] "has in it the heart of quantum mechanics." The simulation program for the double-slit experiment provides an interactive environment for exploring its features in a variety of situations, including the measurement process with the Feynman light source.

2. The second key experiment is the experiment of Grangier, Roger and Aspect [14], which is illustrated in figure 2. Single photons are investigated alternatively at a single beam splitter or in a Mach-Zehnder interferometer with two beam splitters. The experiment consists of two parts: (a) If beam splitter 2 (shown in figure 2 with dashed lines) is not present, the photons encounter a single beam splitter. After passing it, they are registered by two detectors. The detectors measure the coincidence of single photons that are reflected or transmitted at beam splitter 1. (b) If beam splitter 2 is present, the experiment is a Mach-Zehnder interferometer where each photon can reach a detector via two paths.

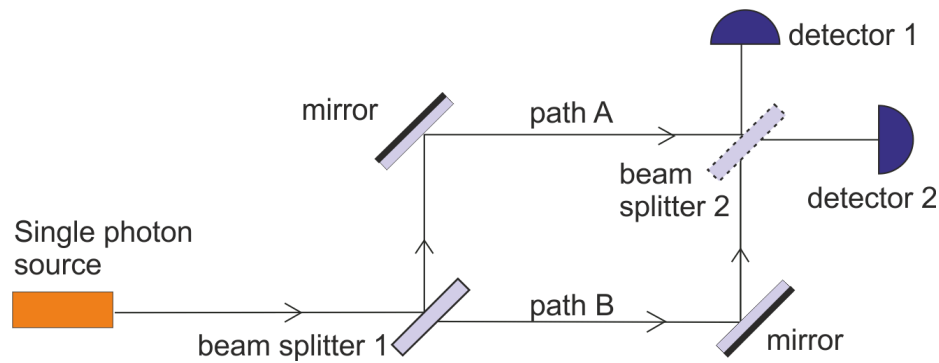

**Fig. 2.** Schematic setup of the experiment by Grangier, Roger and Aspect (1986). In the milq simulation software, the two detectors are replaced by detection screens and the beam is widened to obtain a spatial interference pattern.

## 6    Application of the reasoning tools: Beam splitter and Mach-Zehnder interferometer

Let us illustrate the application of the reasoning tools with the second key experiment:

**a) Rule 3:** Without beam splitter 2, the two detectors 1 and 2 perform a position or path measurement. According to rule 3, each position measurement has a definite result. Exactly one of the two detectors clicks, never both. The two detectors show perfect anticoincidence. On measurement, a single photon is always found at a definite position, never at two positions at the same time. The experiment by Grangier, Roger and Aspect received considerable attention because this anticoincidence provides genuine evidence for the quantum nature of light. The result cannot easily be explained by semiclassical models.

**b) Rule 1:** There is no means to predict at which of the two detectors the next photon will be detected. No physical feature exists that determines whether a photon is transmitted or reflected at the beam splitter. However, a statistical prediction can be made: If the experiment is repeated very often, about half of the photons are found at detector 1 and half at detector 2.

If the second beam splitter is present, rule 1 can be applied too: when a photon hits the screen, its energy is released locally at a specific spot. At which position the next spot is detected cannot be predicted. However, the distribution that is formed when many spots are detected is reproducible: it is the interference pattern known from classical optics (figure 3).

**c) Rule 2:** If beam splitter 2 is inserted, there are two classical alternatives for the experimental result "detector 1 clicks": a photon may have gotten there via path A or via path B. According to rule 2, interference occurs when the path length of the two interferometer arms is varied. This has indeed been observed in the experiment of Grangier, Roger and Aspect. The simulation program is based on a somewhat different mechanism for interference. It is assumed that the two paths are slightly different in length and the beam is widened by a lens. This leads to an interference pattern with a ring-like structure, as shown in figure 3.

**d) Rule 4:** To demonstrate rule 4, the experiment shown in figure 3 can be extended by polarization filters in both arms. Depending on their relative orientation, they imprint path information on the photons through polarization. If both polarization filters are parallel, no which-path information is encoded into the polarization degree of freedom and the interference pattern is found on the screen. However, with polarization filters set perpendicular to each other, which-path information is encoded and no interference pattern appears (figure 4). Which-path information and interference pattern are mutually exclusive – an example of complementary quantities.

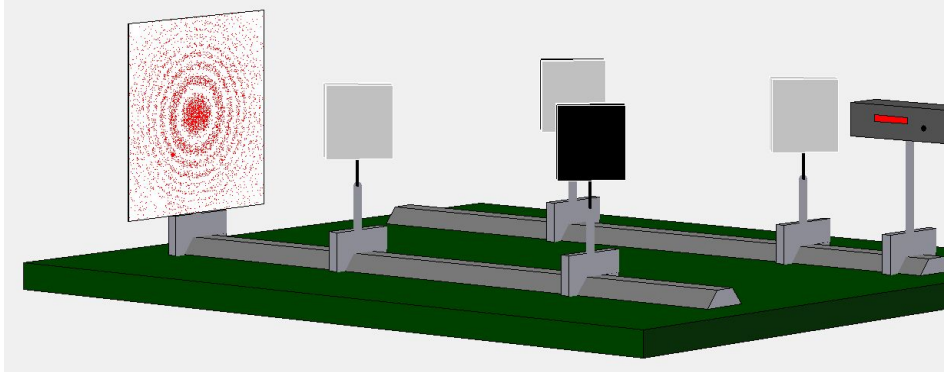

**Fig. 3.** The interference pattern known from classical optics emerges from the detection traces of many single photons. Screenshot of the milq Mach-Zehnder simulation program.

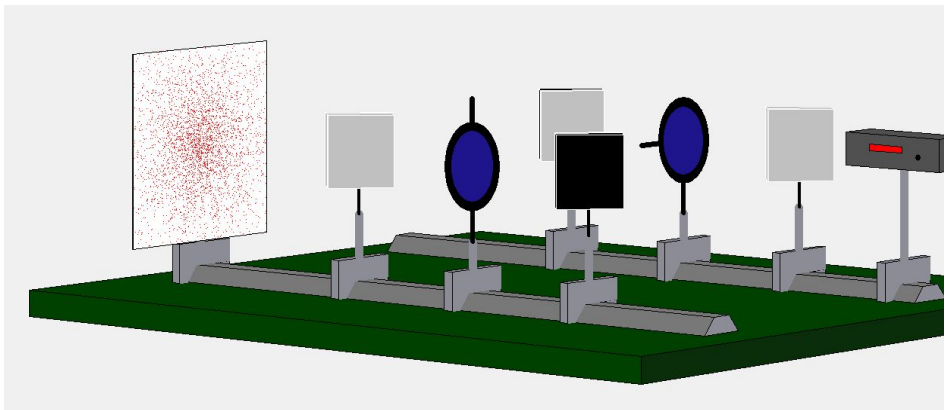

**Fig. 4.** Which-path information is encoded with polarization filters. No interference pattern appears.

## 7   Curriculum

The detailed curriculum of the milq course has already been discussed in [11]. We will therefore only give a brief outline. The course consists of two main parts which form a spiral curriculum (figure 5). First, the quantum physical behavior of photons is examined, using the Mach-Zehnder simulation software. It is shown that light may show both wave and particle aspects in the same experiment. However, none of our classical models is sufficient to describe the phenomena adequately. It is also shown that in an interferometer, a photon cannot be regarded as a particle-like entity which is localized in one of the interferometer arms. Born's probability interpretation is necessary to explain the experimental results. In the second part electrons are studied (mainly in the context of the double-slit experiment). The same topics as with photons are treated, but on a conceptually higher level. The probability interpretation is formulated with the wave function. Still the discussion remains qualitative, i.e. the wave function is only considered as an abstract entity. Finally, some quite advanced topics are discussed: the quantum mechanical measurement process, the paradox of Schrödinger's cat and its resolution by decoherence.

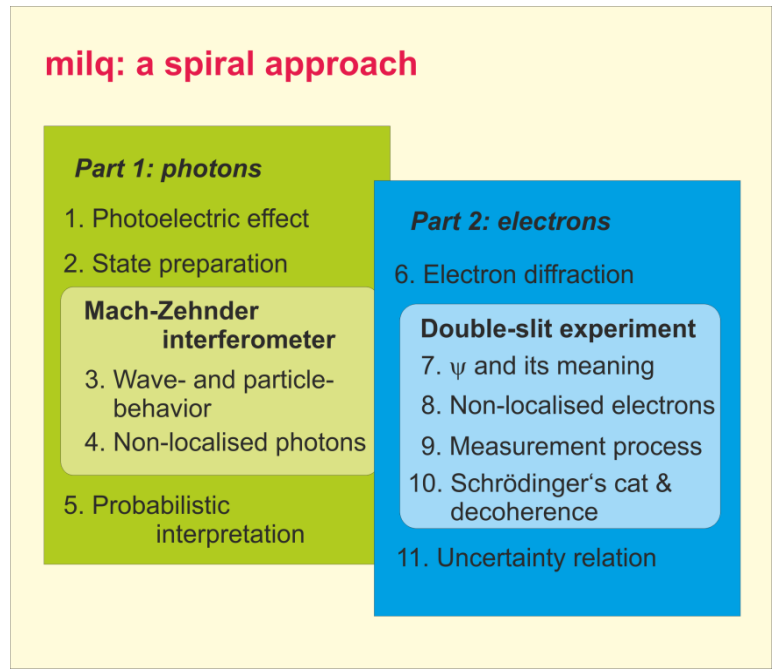

**Fig. 5.** The milq curriculum

## 8    Empirical results

The milq course has been evaluated at the time of its conception. The results are fully documented in [12] (in German); some of the results have been published already in [11]. Here we only give a brief overview. In the evaluation, a mix of qualitative and quantitative methods was used. To assess whether students at high school level are able to deal with the challenging subject matter, the following studies were conducted with high school students in grade 13 (more precisely, "high school" means German Gymnasium which is visited by about half the students of a cohort, one third of them chooses physics):

- teaching experiments with individual students (N = 8),
- semi-structured interviews on student conceptions (N = 23),
- questionnaire on student conceptions (N = 60),
- results of written and oral exams.

**Table 1.** Evaluation results for questionnaire on students' conceptions questionnaire. Effect size is Cohen's *d*; asterisks denote the level of significance (*: $p < 0.05$; **: $p < 0.01$; ***: $p < 0.001$)

| Student conceptions on: | Experimental group (high school students) | Control group (university students) | Effect size $d$ |
|---|---|---|---|
| Structure of the atom (6 items) | +60.9 | +40.8 | 0.65** |
| Determinism/indeterminism (9 items) | +51.6 | +37.4 | 0.47* |
| Dynamical properties of quantum objects (3 items) | +71.6 | +41.9 | 0.83*** |
| Uncertainty relation (10 items) | +51.5 | +30.2 | 0.92*** |
| Overall index (29 items) | +55.8 | +35.2 | 0.97*** |

The questionnaire on students conception contained questions with open answers as well as statements that the students had to rate on a five-point Likert scale from 1 (strongly agree) to 5 (strongly disagree). Four indices were formed from the total of 29 items in the questionnaire, reflecting the students' conceptions on the structure of the atom, determinism/indeterminism, quantum mechanical properties, and the uncertainty relation. In addition, an overall index was calculated by averaging over all 29 items. Both the overall index and the sub-indices were scaled so that the value +100 stands for completely adequate quantum mechanical concepts, while −100 means completely inadequate concepts. The results are given in Table 1. For the high school students, the mean value of the overall index was +55.8 with a standard deviation of 19.5. Such a high value can be interpreted as an indication that the students have successfully built up the desired quantum mechanical concepts through the milq course.

It was difficult to define a suitable control group because the content of the milq course was far removed from the standard high school curriculum. We decided to compare the high school students of the experimental group with a group consisting of 35 first-year physics students from the University of Munich which had not yet attended quantum physics courses in university. They formed a "group of learners with good knowledge of physics" where at least below-average performance was not to be expected. The conceptions of this group were compared with those of the experimental group. The mean value of the overall conception index in this comparison group was +35.2, i.e. 20.6 points lower than in the test group. The difference is statistically highly significant (p < 0.1%). The effect strength d is 0.97; it is therefore a large effect. The differences in the individual areas are also consistently significant to highly significant with medium to large effect sizes (Table 1).

## 9 Application on a European scale

Until 2018, the teaching material for the milq course was only available in German. This made it difficult to use the milq concept outside the German-speaking world. In 2018, a first attempt was made to promote the milq approach in Italy. The material was used as the basis for teacher in-service training at the University of Trieste. Italy is currently in the process of changing the school curricula in physics by adding modern physics. Teachers are being offered structured training to prepare them for the new teaching [15]. A number of teachers approached the University of Trieste and asked for an ad hoc seminar to adapt their knowledge to their teaching needs. The topics of the seminars were chosen based on the requests of the participants after they were introduced to the four quantum reasoning rules. The following sequence of seminar has emerged:

1. Introduction of the "quantum reasoning tools",
2. the photoelectric effect,
3. single photon in a Mach-Zehnder interferometer,
4. measurement of Planck's constant using LED,
5. construction of a spectrometer used to measure a Planck's constant with LEDs.

Although there was no systematic evaluation, the feedback from the teachers was positive. They found the work with the reasoning tools useful and were interested in continuing the training. The activity will be continued and published in the near future.